\documentclass{IPEC2026}
\IEEEoverridecommandlockouts

\usepackage{hyperref}
\hypersetup{hidelinks}
\usepackage{xcolor,soul,framed} 

\colorlet{shadecolor}{yellow}
\usepackage[pdftex]{graphicx}
\DeclareGraphicsExtensions{.pdf,.jpeg,.png, .jpg }

\usepackage[table]{xcolor}
\definecolor{light-gray}{gray}{0.9} 
\usepackage{array}   

\usepackage{stfloats}
\usepackage{amsfonts,amssymb}
\usepackage{mathrsfs}
\usepackage{amsthm}
\usepackage[cmex10]{amsmath}
\usepackage{mathtools}
\usepackage{bbm}
\DeclareMathAlphabet{\mathbbb}{U}{bbold}{m}{n}
\usepackage{array}
\usepackage{flushend}
\usepackage{physics}
\usepackage{cite}
\usepackage{comment}
\usepackage{graphicx}
\usepackage[american,cuteinductors,smartlabels]{circuitikz}
\theoremstyle{remark}
\newtheorem{theorem}{Theorem}

\newcommand{\ie}{i.e.}

\usepackage{amsmath,amssymb}

\def\BibTeX{{\rm B\kern-.05em{\sc i\kern-.025em b}\kern-.08em
    T\kern-.1667em\lower.7ex\hbox{E}\kern-.125emX}}
\begin{document}

\title{ A Unified Framework for Contraction Stability Analysis of Heterogeneous Grid-Forming Inverters 
}

\track{: 1 or 2,
Qianxi~Tang,~\IEEEmembership{Student Member,~IEEE,}
\ and~Li~Peng,~\IEEEmembership{Senior Member,~IEEE}%
\thanks{This paper has not been presented elsewhere previously. The authors are with School of Electrical and Electronic Engineering, Huazhong University of Science and Technology, Wuhan 430074, China. Email: qianxi@hust.edu.cn, pe105@mail.hust.edu.cn}}

\maketitle

\begin{abstract}
The shift to renewable-dominated power systems has produced low-inertia grids, undermining system stability. In this context, grid-forming inverter (GFM) become a solution.
 However, GFMs challenge conventional analysis techniques, especially those relying on small-signal or root-mean-square (RMS) models. Small-signal and RMS models rely on linearization and sinusoidal steady state. In large-signal cases, these assumptions fail.  Stability of GFM-based system becomes operating-point dependent, and a feasible operating point may not even exist. While large-signal analyses are available, decentralized certification of operating-point convergence with explicit transient guarantees (e.g., rate, overshoot) remains rare. This paper proposes an algebraic, decentralized contraction-based framework. The contraction stability analysis in the framework certifies the system stability and convergence to desired operating points. The method works in the time domain and captures nonlinear, large-signal behavior of synchronization and power-sharing mechanisms. Moreover, the contraction rate in the proposed framework provides an explicit bound on transient time—trajectories converge exponentially to the new operating point at a controlled rate—yielding computable contraction regions that certify stability and large-signal convergence across operating-point changes. These regions directly guide parameter tuning for heterogeneous GFMs.
	
\end{abstract}

\begin{IEEEkeywords}
	Synchronization, grid-forming inverters,  transient stability.
\end{IEEEkeywords}

\section{Introduction}
\IEEEPARstart{T}{he} global shift toward nearly 100\% renewable generation is displacing synchronous machines—and with them inherent inertia, voltage support, and frequency regulation—leaving low-inertia grids dominated by power-electronic converters \cite{b0,b1,b2}. Despite the promise of grid-forming inverter strategies, guaranteeing stability for these nonlinear, time-varying, converter-dominated systems remains a central open challenge \cite{b3}.
Existing approaches to stability analysis expose the limits of traditional methods. Small-signal \cite{b4} and quasi-static RMS models \cite{b5}—widely used to characterize the dynamics of synchronous machines and early converter systems—fall short when applied to modern converter-dominated grids. For instance, models based on RMS or phasor representations, which treat large time-varying dynamics as algebraic manipulations of complex numbers, fail to capture critical instability phenomena, such as those induced by phase-locked loops (PLLs), as highlighted in \cite{b6}. This casts fundamental doubt on the validity of quasi-steady-state analysis techniques in evaluating converter-driven systems. To address shortcomings of small-signal \cite{b4} and quasi-static analyses, researchers have used Lyapunov theory \cite{b1}, passivity-based methods \cite{b7}, and RMS infinite-series expansions \cite{b8} for understanding large-signal/transient stability; however, decentralized guarantees of operating-point convergence with explicit transient metrics (rate, overshoot) are rare.

What is needed is a unified, scalable theory of stability for converter-dominated grids—one that can rigorously analyze nonlinear and time-varying behavior, accommodate large disturbances, and offer tractable insight into the interactions of diverse devices and control strategies.

Contraction analysis offers a powerful and underutilized alternative. Instead of focusing on the stability of a particular equilibrium point, contraction theory asks whether the distance between trajectories decreases over time—indicating that all trajectories converge toward one another, independent of initial conditions or temporary disturbances \cite{b9}. This shift enables global and incremental stability analysis even for systems with nonlinear, time-varying, or large-signal behavior. Critically, contraction analysis operates directly in the time domain and avoids the assumption of algebraic complex-number linearity inherent in RMS and quasi-static phasor models. This makes it fundamentally better suited to capturing the true dynamics of modern power systems, especially where phase and frequency interactions dominate. Furthermore, contraction permits decentralized stability certification through local Jacobian and metric-based conditions—making it scalable for system-level integration of heterogeneous inverters. It has already proven useful in related contexts, such as voltage synchronization and current sharing in inverter networks, but has not yet been developed into a general framework for analyzing heterogeneous GFM-connected grids.

This paper proposes an algebraic, decentralized contraction-based framework for analyzing the stability of GFM-based power systems. This novel method avoids equilibrium-method fragility and RMS inaccuracies, and provides explicit transient-time bounds (exponential convergence at a set rate) with computable contraction regions that certify large-signal stability and directly guide parameter tuning for heterogeneous GFMs.
Classical droop, complex droop, also called dispatchable virtual oscillator control (dVOC), and virtual synchronous machine (VSM) controls with a common RL interconnection in the IEEE 9-bus test system will be analyzed by this unified, decentralized framework for the GFM-based power system.

\section{Contraction Stability Analysis for Grid-forming Inverters}
\label{sec:passivity analysis}


In this section, the contraction concepts will be illustrated with some corresponding theorems. It can be shown that even with modeling error or deterministic disturbance, the contraction guarantees all the trajectories of grid-forming inverters converge to a particular solution (static-state) exponentially with a bounded steady-state error.

\begin{figure}
	\begin{center}
		\includegraphics[width = 1\linewidth]{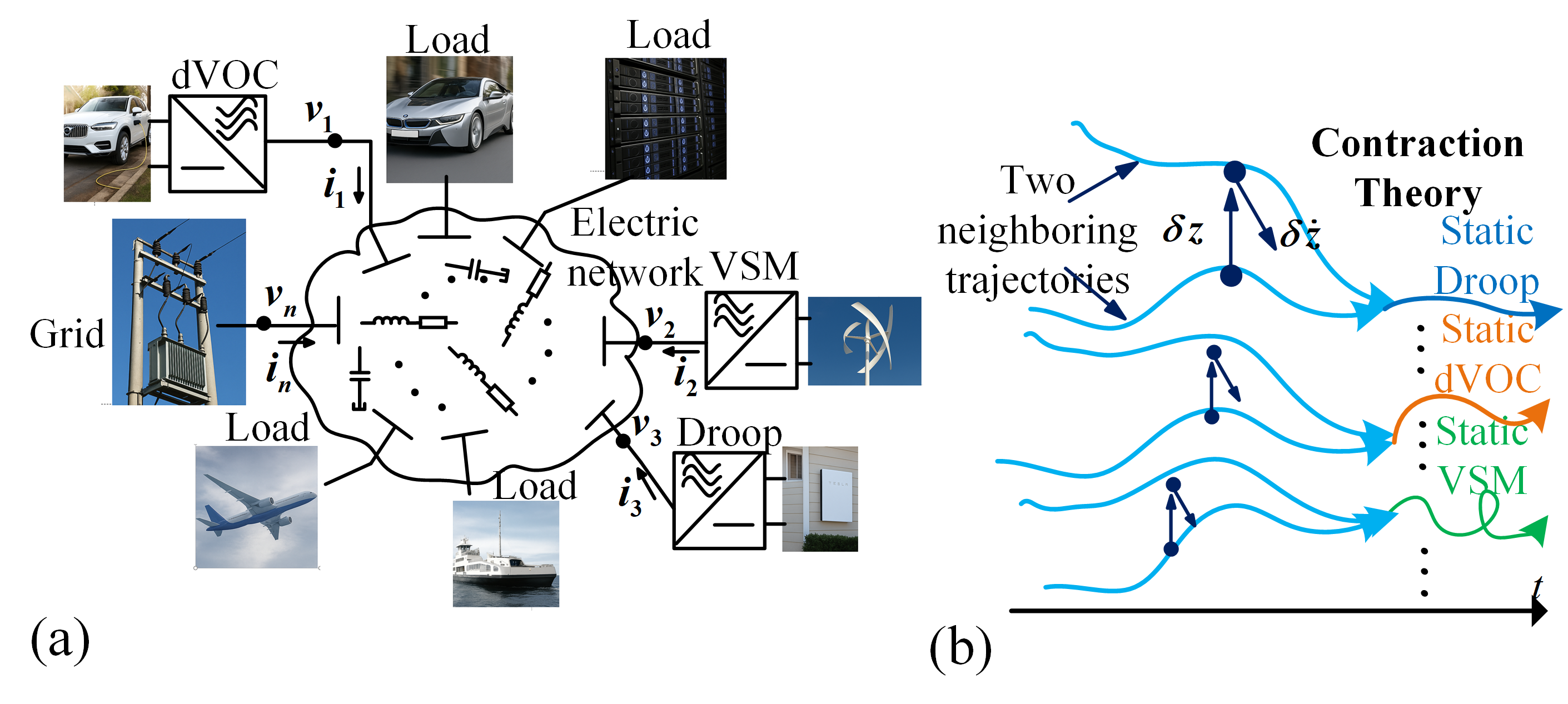}
		\caption{ The studied GFM-based power system diagram with the proposed contraction-based stability analysis. (a) A general GFM-based power system diagram. (b) The diagram of contracting trajectories.}
		\label{fig:block-diagram}
	\end{center}
\end{figure}

\subsection{Preliminaries and Notation}
\label{notation}
For a square matrix $M^{n\times n}$, $M \succ 0$, $M \succeq 0$, $M \prec 0$, and $M \preceq 0$ are for the positive definite, positive semi-definite, negative definite, negative semi-definite matrices, respectively. $x_i$ and $x_j$ are the $i$th and $j$th elements of ${\mathbf{x}} \in \mathbb{R}^n$. To describe partial derivatives, $f_{{\mathbf{x}}} = \partial f/\partial {\mathbf{x}}$, $M_{x_i} = \partial M/\partial x_i$, and $M_{x_ix_j} = \partial^2 M/(\partial x_i\partial x_j)$ are used. The other notations are given in Table~\ref{tab:notations_in_this_paper}. Let $\mathbf v=\begin{bmatrix}v_\alpha & v_\beta\end{bmatrix}^\top$ and $\mathbf i=\begin{bmatrix}i_\alpha & i_\beta\end{bmatrix}^\top$ denote the inverter voltage and filter current in stationary $\alpha\beta$ coordinates. We define $r=\|\mathbf v\|^2=v_\alpha^2+v_\beta^2$, $v=\sqrt r$, and the $90^\circ$ rotation matrix $J=\begin{bmatrix}0 & -1\\ 1 & 0\end{bmatrix}$. The active and reactive powers are then given by $p=\mathbf v^\top\mathbf i$ and $q=(J\mathbf v)^\top \mathbf i$. The filter dynamics connecting the inverter to the grid are described by $L_f\dot{\mathbf i}=(\mathbf v-\mathbf v_o)-R_f\mathbf i$, where $\mathbf v_o(t)$ is the PCC voltage, here treated as a time-dependent variable whose role will be clarified later. A key identity used throughout is the kinematic decomposition $\dot{\mathbf v}=S(\cdot)\mathbf v+\Omega(\cdot)J\mathbf v$, where $S=\dot v/v$ and $\Omega=\dot\theta$, which holds for any differentiable voltage trajectory $\mathbf v(t)$.

\begin{table}
	\caption{Notations used in this paper. \label{tab:notations_in_this_paper}}
	\footnotesize
	\begin{center}
		\renewcommand{\arraystretch}{1.2}
		\begin{tabular}{ |c|m{6.3cm}| } 
			\hline
			$\|{\mathbf{x}}\|$ & Euclidean norm of $x \in \mathbb{R}^n$ \\ \hline
			$\delta {\mathbf{x}}$ & Differential displacement of $x \in \mathbb{R}^n$ \\ \hline
			$\|M\|$ & Induced $2$-norm of $M \in \mathbb{R}^{n\times m}$ \\ \hline
			$\lambda_{\min}(M)$  & Minimum eigenvalue of $M\in\mathbb{R}^{n\times n}$ \\ \hline
			$\lambda_{\max}(M)$  & Maximum eigenvalue of $M\in\mathbb{R}^{n\times n}$ \\ \hline $\mathrm{I}$ & Identity matrix of appropriate dimensions \\ \hline
			$\mathbb{R}_{>0}$ & Set of positive reals, \ie{}, $\{a\in\mathbb{R}|a\in(0,\infty)\}$ \\ \hline
			$\mathbb{R}_{\geq 0}$ & Set of non-negative reals, \ie{}, $\{a\in\mathbb{R}|a\in[0,\infty)\}$ \\ \hline
		\end{tabular}
	\end{center}
\end{table}

\subsection{Contraction Theorems }

Consider the smooth non-autonomous system
\begin{equation}
	\label{eq:xfx}
	\dot{\mathbf{x}}(t)=f\bigl(\mathbf{x}(t),t\bigr),
\end{equation}
where $t \in \mathbb{R}_{\geq 0}\; \text{is time}, \text{states}\; \mathbf{x}:\mathbb{R}_{\ge0}\!\to\mathbb{R}^n,\;
f:\mathbb{R}^n\times\mathbb{R}_{\ge0}\!\to\mathbb{R}^n$
whose smoothness ensures local existence and uniqueness 
of the solution to \eqref{eq:xfx} for a given $\mathbf{x}(0)=\mathbf{x}_0$ at least locally. 

\begin{theorem}[Contracting]
	\label{Thm:contraction}
	If there exists a uniformly positive definite matrix ${M}({\mathbf{x}},t)={{\Theta}}({\mathbf{x}},t)^{\top}{{\Theta}}({\mathbf{x}},t) \succ 0,~\forall \mathbf{x},t$, where ${\Theta(\mathbf{x},t)}$ defines a smooth coordinate transformation of $\delta x$, \ie, $\delta{z}={\Theta}(\mathbf{x},t)\delta{\mathbf{x}}$, \st{} either of the following equivalent conditions holds for $\exists \alpha \in \mathbb{R}_{>0}$, $\forall \mathbf{x},t$:
	\begin{align}
		&\lambda_{\max}(F(\mathbf{x},t))=\lambda_{\max}\left(\left({\dot{\Theta}}+{{\Theta}}\frac{\partial
			{f}}{\partial
			{\mathbf{x}}}\right){{\Theta}}^{-1}\right) \leq - \alpha
		\label{eq_MdotContracting_z} \\
		&{\dot{M}}+M\frac{\partial
			{f}}{\partial {{\mathbf{x}}}}+\frac{\partial {f}}{\partial
			{\mathbf{x}}}^{\top}M \preceq -2\alpha M,
		\label{eq_MdotContracting}
	\end{align}
	where the arguments $(\mathbf{x},t)$ of $M(\mathbf{x},t)$ and $\Theta(\mathbf{x},t)$ are omitted for notational simplicity, then all the solution trajectories of \eqref{eq:xfx} converge to a single trajectory exponentially fast regardless of their initial conditions (\ie{}, contracting), with an exponential convergence rate $\alpha$~\cite{b9}. The converse also holds.
\end{theorem}
\begin{theorem}[Robustness under perturbation]
	\label{thm:incstab_robust}
	Consider $\dot {\mathbf{x}} = f( {\mathbf{x}},t)$ and suppose there exists a smooth Riemannian metric
	$M( {\mathbf{x}},t)=\Theta({\mathbf{x}},t)^{\!\top}\Theta({\mathbf{x}},t)\succ 0$ and $\alpha>0$ such that
	\begin{equation}
		\dot M + M\frac{\partial f}{\partial \mathbf{x}} + \frac{\partial f}{\partial {\mathbf{x}}}^{\!\top} M \;\preceq\; -\,2\alpha\,M
		\quad \text{for all }({\mathbf{x}},t).
		\label{eq:contr_cond}
	\end{equation}
	Let $\xi_0(t)$ and $\xi_1(t)$ be any two solutions of $	\dot{\mathbf{x}}(t)=f\bigl(\mathbf{x}(t),t\bigr)$ and let
	\[
	V_\ell(t) \;:=\; \inf_{\text{paths }{\mathbf{x}}(\mu,t)} \int_{0}^{1} \bigl\|\Theta({\mathbf{x}},t)\,\partial_\mu x\bigr\|\,d\mu
	\]
	be the Riemannian path length (geodesic distance) between $\xi_0(t)$ and $\xi_1(t)$.
	Then
	\begin{equation}
		V_\ell(t) \;\le\; e^{-\alpha t}\,V_\ell(0),
		\qquad
		\|\xi_1(t)-\xi_0(t)\| \;\le\; \frac{e^{-\alpha t}}{\sqrt{\underline m}}\,V_\ell(0),
		\label{eq:inc_exp_bound}
	\end{equation}
	whenever $M({\mathbf{x}},t)\succeq \underline m I$~\cite{b9}.
	Hence, any two trajectories converge exponentially (incremental stability).
	Now consider the perturbed system $\dot {\mathbf{x}}=f({\mathbf{x}},t)+d({\mathbf{x}},t)$ with $\|d(x,t)\|\le\bar d$ and
	assume $\underline m I \preceq M({\mathbf{x}},t)\preceq \overline m I$.
	Then the geodesic distance and the Euclidean separation satisfy~\cite{b9}
	\begin{align*}
		V_\ell(t)
		&\le e^{-\alpha t}V_\ell(0)\\
		&+ \frac{\sup_{{\mathbf{x}},t}\|\Theta({\mathbf{x}},t)\,d({\mathbf{x}},t)\|}{\alpha}\,\bigl(1-e^{-\alpha t}\bigr), \\
		\|\xi_1(t)-\xi_0(t)\|
		&\le \frac{V_\ell(0)}{\sqrt{\underline m}}\,e^{-\alpha t}
		+ \frac{\bar d}{\alpha}\sqrt{\frac{\overline m}{\underline m}}\,(1-e^{-\alpha t}).
		\label{eq:robust_ball}
	\end{align*}
	
	Consequently, the separation decays exponentially and remains bounded by a disturbance-to-state “error ball” of radius
	$\displaystyle \frac{\bar d}{\alpha}\sqrt{\tfrac{\overline m}{\underline m}}$.
\end{theorem}


\section{Plant, Controllers, and Unified Models}\label{sec:models}

\subsection{Classical Droop (polar coordinates to $\alpha\beta$ coordinates)}
\paragraph{Polar (voltage magnitude and angle)~\cite{b3}}
\begin{equation}
	\dot v=\eta\,(q^\star-q)+\eta\alpha\,(v^\star-v),\,
	\dot\theta=\omega_0+\eta\,(p^\star-p).
\end{equation}

\paragraph{$\alpha\beta$ form and kinematic scalars}
Using $p=\mathbf v^\top\mathbf i$, $q=(J\mathbf v)^\top\mathbf i$, we get
\begin{equation}\label{eq:S_Omega_class}
	S=\frac{\eta\,(q^\star-q)+\eta\alpha\,(v^\star-v)}{v},\,
	\Omega=\omega_0+\eta\,(p^\star-p).
\end{equation}

\subsection{dVOC (polar coordinates to $\alpha\beta$ coordinates)}
\paragraph{Polar~\cite{b3}}

\[
\frac{\dot v}{v} =  
\eta\!\left( \frac{q^\star}{(v^\star)^2} - \frac{q}{r} \right)
+ \eta\alpha\,\frac{v^\star - v}{v^\star}
,\,
\dot\theta = \omega_0 + \eta\!\left( \frac{p^\star}{(v^\star)^2} - \frac{p}{r} \right).
\]

\paragraph{$\alpha\beta$ kinematic scalars}

\[
S=\eta\!\left(\frac{q^\star}{(v^\star)^2}-\frac{q}{r}\right)+\eta\alpha\,\frac{v^\star-v}{v^\star},\,
\Omega=\omega_0+\eta\!\left(\frac{p^\star}{(v^\star)^2}-\frac{p}{r}\right).
\]

\subsection{VSM~\cite{b2}}
Let the internal electromagnetic force be aligned with $\mathbf v/v$ and magnitude $E$.
Under a fast field loop (typical in VSM), $E\simeq v^\star$ is regulated quasi-constantly,
so $S\equiv 0$ and
\begin{equation}\label{eq:vsm_kin}
	\dot{\mathbf v}=\Omega\,J\mathbf v,\qquad
	J_r\,\dot\omega=\frac{1}{\omega^\star}(p^\star-p)+D_p(\omega^\star-\omega),
\end{equation}
with $\Omega\equiv\omega$ and $D_p>0$.
(If field dynamics are retained, $S=\dot E/E$ couples to the PI loop.)
Here, $S\equiv 0$, $\Omega=\omega$ (state).
The expected power sharing dynamics (the static state) appears when the derivatives of voltage magnitude and angle are zero.


\section{A Novel Unified Contraction-based Framework for GFM-based Power System}

\subsection{Contraction Stability Criterion Using General Symmetric Jacobian and Schur Test}
Define the full state $x=[\mathbf v;\mathbf i;\omega]$ for the Jacobian matrix. Let $\mathrm{Sym}(M):=\tfrac12(M+M^\top)$. Partition the symmetric Jacobian as
\begin{equation}
	\mathrm{Sym}(J)=
	\begin{bmatrix}
		A & B & B_\omega\\
		B^\top & D & 0\\
		B_\omega^\top & 0 & D_\omega
	\end{bmatrix},
\end{equation}

In the Euclidean metric the (local) contraction rate is
\[
c(x)\ :=\ -\,\lambda_{\max}\!\big(\mathrm{Sym}(J)(x)\big),
\]
so we need an \emph{upper} bound on $\lambda_{\max}(\mathrm{Sym}(J))$.

Since $D\prec 0$ (and $D_\omega<0$ when a swing loop is present), contraction in the
Euclidean metric follows from the Schur complements:
\begin{equation}\label{eq:schur}
	A_{\rm eff}:=
	A-\underbrace{B\,D^{-1}B^\top}_{\frac{L_f}{R_f}BB^\top}
	-\underbrace{B_\omega\,D_\omega^{-1}B_\omega^\top}_{\frac{J_r}{D_p}B_\omega B_\omega^\top\ \text{(VSM)}}\ \prec\ 0.
\end{equation}

From eigenvalues to a rate bound via the Schur complement, a conservative local contraction rate is
\begin{equation}\label{eq:crate_general}
	c(x)\ \ge\ \min\!\left\{\frac{R_f}{L_f},\ \frac{D_p}{J_r},\ -\lambda_{\max}\big(A_{\rm eff}\big)\right\},
\end{equation}
(with the $D_p/J_r$ term omitted for droop/dVOC without an explicit swing state).

\begin{figure}[!h]
	\centering
	\tikzstyle{comp} = [coordinate]
	\makeatletter
	\begin{circuitikz}[scale=1.2,line width=0.1mm,rounded corners=0mm]
		{\fontfamily{ptm}\selectfont    
			\draw(0,1) circle(0.25cm);
			\draw(0.25,1)to[short]++(0.25,0);
			\draw(0.5,1)to[voosource]++(1,0);            
			\begin{scope}[line width=0.175mm]
				\draw (0.5,0.75)--node[pos=1,above]{\scriptsize 1}(0.5,1.25);
				\draw (1.5,0.75)--node[pos=1,above]{\scriptsize 4}(1.5,1.25);
				\draw (2.25,1.25)--node[pos=0.5,above]{\scriptsize 9}(2.75,1.25);
				\draw (1.5,-0.25)--node[pos=1,above]{\scriptsize 5}(1.5,0.25);
				\draw (2.25,-0.25)--node[pos=0.5,above]{\scriptsize 6}(2.75,-0.25);
				\draw (2.25,-1.25)--node[pos=1,right]{\scriptsize 3}(2.75,-1.25);
				\draw (3.5,0.75)--node[pos=1,above]{\scriptsize 8}(3.5,1.25);
				\draw (3.5,-0.25)--node[pos=1,above]{\scriptsize 7}(3.5,0.25);
				\draw (4.5,0.75)--node[pos=1,above]{\scriptsize 2}(4.5,1.25);
				\node[] at (0,1) {\scriptsize  Droop};
				\node[] at (5,1) {{\scriptsize dVOC}};	
				\node[] at (2.5,-1.75) {\scriptsize VSM};    
			\end{scope}            
			\draw (1.5,1.1)-|(2.4,1.25);
			\draw (1.5,0.9)--(1.65,0.9)--(1.65,0.1)--(1.5,0.1);
			\draw (1.5,-0.1)-|(2.4,-0.25);
			\draw (2.6,-0.25)|-(3.5,-0.1);
			\draw (3.5,0.1)--(3.35,0.1)--(3.35,0.9)--(3.5,0.9);
			\draw (3.5,1.1)-|(2.6,1.25);
			\draw (2.5,1.25)--(2.5,0.75)--(2.75,0.75)--(2.5,0.45)--(2.25,0.75)--(2.5,0.75);
			\draw (1.5,0)--(1.25,0)--(1.25,-0.5)--(1.5,-0.5)--(1.25,-0.8)--(1,-0.5)--(1.25,-0.5);
			\draw (3.5,0)--(3.75,0)--(3.75,-0.5)--(4,-0.5)--(3.75,-0.8)--(3.5,-0.5)--(3.75,-0.5);            
			\draw(4.75,0.75) rectangle (5.25,1.25);
			\draw(4.75,1)to[short]++(-0.25,0);
			\draw(4.5,1)to[voosource]++(-1,0);
			\draw(2.5,-1.25)to[voosource]++(0,1);
			\draw(2.25,-2) rectangle (2.75,-1.5);
			\draw(2.5,-1.5)to[short]++(0,0.25);            
		}
	\end{circuitikz}
	\caption{IEEE 9-bus test system with aggregated grid-forming inverters (VSM, droop, dVOC) and constant-impedance loads. \label{9bus_system}} 
	\label{Test}
\end{figure}
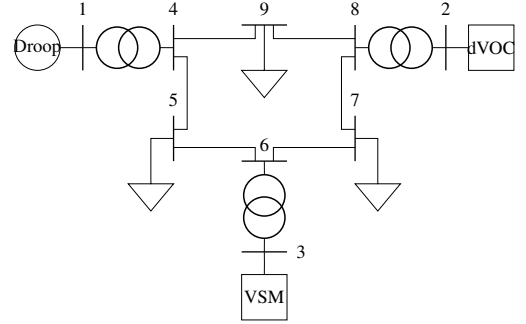

\subsection{Application of Contraction Stability Criterion and Contraction Rates for Specific Controllers}

\paragraph{Classical droop:} In this case the cross Jacobian is 
$J_{vi}=\eta\!\left(\tfrac{1}{v}\,\mathbf v(J\mathbf v)^\top-(J\mathbf v)\mathbf v^\top\right)$, 
which is skew-symmetric. Consequently one has $BB^\top\succeq \tfrac{1}{4L_f^2}I_2$, and a sufficient contraction condition is obtained as $A-\tfrac{1}{4R_fL_f}I_2\prec 0$.

\paragraph{dVOC:} For dispatchable VOC, $J_{vi}=\eta J$ is constant and skew, so $BB^\top=\tfrac14(\eta^2+1/L_f^2)I_2$ and the effective Jacobian reduces to $A_{\rm eff}=A-\Delta_{\rm dVOC}I_2$ with $\Delta_{\rm dVOC}=\tfrac{L_f}{R_f}(\eta^2+1/L_f^2)/4$.

\paragraph{VSM (fast field):} When $S\equiv 0$, the block $A$ vanishes and only the Schur-complement terms remain, giving a guaranteed contraction rate of at least $\min\{R_f/L_f,\;D_p/J_r\}$ uniformly over the electrical phase.

For droop and dVOC the spectral bound can be simplified. A two–by–two estimate yields $\lambda_{\max}(A)\le |S|+\tfrac{v}{2}(\|\nabla_{\!v}S\|+\|\nabla_{\!v}\Omega\|)$. Using the bounds $|p|\le v\|\mathbf i\|$, $|q|\le v\|\mathbf i\|$, and $r=v^2$, and closing with a current envelope at grid frequency $\omega_g$, the current satisfies 
$\|\mathbf i\|\le I_{\max}(R_f,L_f)$ with 
$I_{\max}(R_f,L_f)=K_{\rm tr}(V_{\max}+V_{o,\max})/\sqrt{R_f^2+(\omega_g L_f)^2}$ 
for a conservative constant $K_{\rm tr}\approx 2.5$. With $v_{\min}\le v\le v_{\max}$ and $\|\mathbf i\|\le I_{\max}$, this gives the bound 
$\lambda_{\max}(A)\le\Lambda(R_f,L_f;\eta,\alpha)$, where 
$\Lambda(R_f,L_f;\eta,\alpha)=\tfrac{4\eta I_{\max}(R_f,L_f)}{v_{\min}}+\eta\alpha\!\left(1+\tfrac{3v_{\max}}{2v^\star}\right)$.

It follows that if voltage and current remain bounded, the contraction rate is determined by the local impedance. Thus the next step is to characterize values of $(R_f,L_f)$ that guarantee contraction. Fixing a target contraction rate $c_0$ and a load-impedance cap $|j\omega_g L_f+R_f|<k R_{\rm load}$, a sufficient feasibility set $\mathcal C(\eta,\alpha)$ is given by the pairs $(R_f,L_f)$ that satisfy simultaneously: (i) $R_f/L_f \ge c_0$, (ii) $1/(4R_fL_f)-\Lambda_{\eta,\alpha}(R_f,L_f)\ge c_0$, and (iii) $\sqrt{R_f^2+(\omega_g L_f)^2}<R_{\rm load}$.

\section{Validation: Contraction Stability Criterion With Contraction Regions}

\begin{figure}[!t]
	\begin{center}
		\includegraphics[width = 1\linewidth]{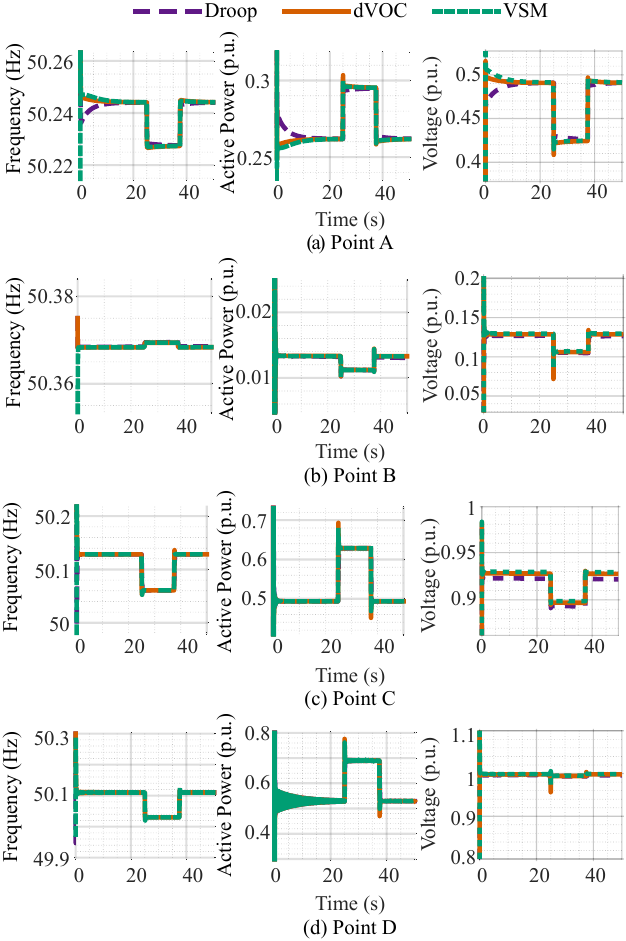}
	\caption{The results of frequency, active power and voltage  of four different $(R_f,L_f)$ parameter settings. }
	\label{fig:points-arch}
	\end{center}
\end{figure}

To illustrate the proposed framework, this section considers a benchmark IEEE 9-bus system shown in Fig.~\ref{Test} whose structure and rated nominal values follow the specifications in \cite{b2}. The test setup consists of converter-interfaced generation connected through local filters to a PCC, which is in turn coupled to the MV/HV network through a transformer stage. The base values, transformer parameters, and control settings are taken directly from \cite{b2}, ensuring consistency with standard case-study practice in the literature.  In both cases, the filter resistance $R_f$ and inductance $L_f$ serve as the key local impedance variables, shaping the damping and dynamic interaction between the inverter and the grid. 
To begin with, the original network load is set to 2.25 p.u., with an additional disturbance of 0.75 p.u. applied at 25 seconds and removed at 37.5 seconds. 
\begin{figure}
	\begin{center}
		\includegraphics[width = 0.95\linewidth]{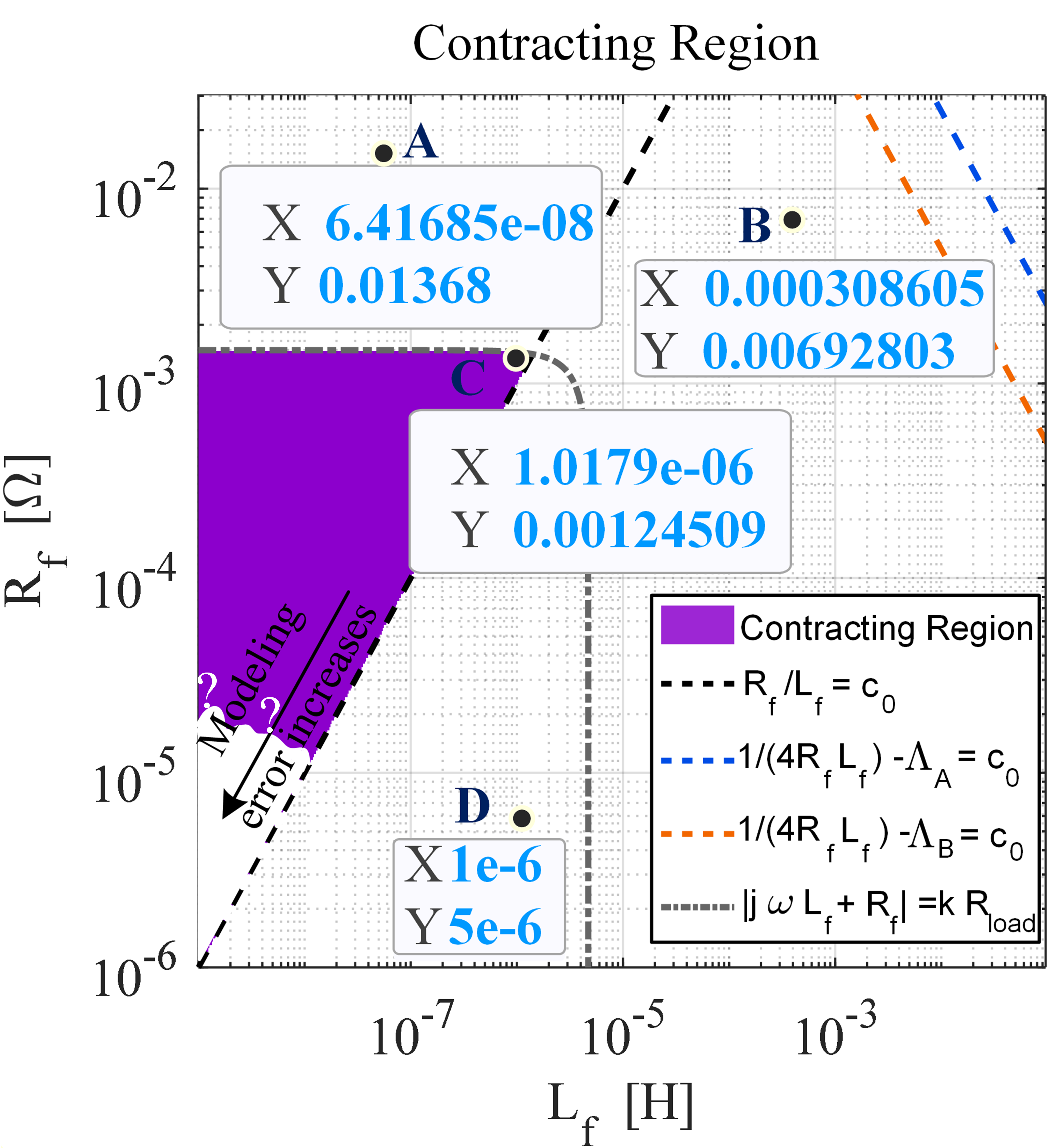}
		\caption{Contraction region for heterogeneous GFMs—certifies the stability domain and quantifies the convergence rate, directly guiding parameter tuning.}
		\label{contractionregion}
	\end{center}
\end{figure}
Four different $(R_f,L_f)$ parameter settings, shown in Fig.~\ref{fig:points-arch}, are considered for comparison, with point D corresponding to the configuration used in \cite{b2}.  It is shown that the point C in the contraction region has a good performance both at the transienceof start-up and the load disturbance.
Shown below are the numerical envelopes used to derive the $\mathcal C(\eta,\alpha)$, defined by the admissible pairs $(R_f,L_f)$ in Fig~\ref{contractionregion}. It takes $v^\star=816.4966$, $v_{\min}=0.2\,v^\star$, $v_{\max}=1.3\,v^\star$,
$\omega_g=2\pi\times 50$, and the conservative $K_{\rm tr}=2.5$.
$k=1/30 $ in $|j\omega_g L_f+R_f|<k R_{\rm load}$,  $c_0=1000 s^{-1}$ with a large contraction to make sure the modeling error effect will be suppressed such as the modeling of $v_o$. The controller parameters for droop are chosen as $\eta_A=3.1416\times 10^{-8}$ same as the $d_w=2\pi*0.05$ of droop in \cite{b2} and $\alpha_A=10^{-3}$ as $k_i=0.001$ of droop in \cite{b2}. Also, the parameters for dVOC are set to $\eta_B=0.0209$ and $\alpha_B=6.6667\times 10^4$ the same as the dVOC in \cite{b2}. The VSM parameters, $D_p=10^5$ and $J=2*10^3, k_p=0.001, k_i=0.0021$, are  the same as in \cite{b2}. Also, the power setpoints of the 3 GFMs are the same. Following the analysis developed in the preceding sections, the feasible regions were given in the $(R_f,L_f)$ plane where the contraction inequalities are satisfied. These regions delineate boundaries between good and potentially unstable parameter choices.

\section{Conclusion}
This paper proposes a unified, decentralized framework to analyze contraction stability for heterogeneous GFMs, such as classical droop, complex droop (dVOC-style), and VSM GFMs. 
Moreover, it derives contraction stability criterion using general symmetric jacobian and schur test that certifies the system stability and convergence to desired operating points with a bounded rate. Additionally, it helps parameter tuning of GFMs by using contraction region in the $(R_f,L_f)$ plane. Within this region, trajectories converge exponentially to desired operating points with bounded error and rate, even under large-signal transients and nonlinear dynamics.
Future work should incorporate detailed network dynamics, digital delays, and non-ideal hardware effects so that the results remain valid under broader and more practical operating conditions.

\end{document}